\documentstyle[12pt,fleqn]{article}

\font\twlgot =eufm10 scaled \magstep1
\font\egtgot =eufm8
\font\sevgot =eufm7
\font\twlmsb =msbm10 scaled \magstep1
\font\egtmsb =msbm8
\font\sevmsb =msbm7

\newfam\gotfam

\textfont\gotfam\twlgot
\scriptfont\gotfam\egtgot
\scriptscriptfont\gotfam\sevgot

\newfam\msbfam
\textfont\msbfam\twlmsb
\scriptfont\msbfam\egtmsb
\scriptscriptfont\msbfam\sevmsb
\def\Bbb{\protect\pBbb}
\def\pBbb{\relax\ifmmode\expandafter\Bb\else\typeout{You cann't use
Bbb in text mode}\fi}
\def\Bb #1{{\fam\msbfam\relax#1}}

\def\thebibliography#1{\section*{References}\list
   {[\arabic{enumi}]}{\settowidth\labelwidth{#1}\leftmargin\labelwidth
     \advance\leftmargin\labelsep
     \usecounter{enumi}}
     \def\newblock{\hskip .11em plus .33em minus .07em}
     \sloppy\clubpenalty4000\widowpenalty4000
     \sfcode`\.=1000\relax}

\let\Large=\large

\def\op#1{\mathop{\fam0 #1}\limits}

\newcommand{\beq}{\begin{equation}}
\newcommand{\eeq}{\end{equation}}
\newcommand{\ben}{\begin{eqnarray}}
\newcommand{\een}{\end{eqnarray}}
\newcommand{\be}{\begin{eqnarray*}}
\newcommand{\ee}{\end{eqnarray*}}
\newcommand{\bea}{\begin{eqalph}}
\newcommand{\eea}{\end{eqalph}}

\newcommand{\al}{\alpha}

\newcommand{\bt}{\beta}
\newcommand{\dl}{\delta}
\newcommand{\la}{\lambda}

\newcommand{\Om}{\Omega}
\newcommand{\m}{\mu}

\newcommand{\vt}{\vartheta}

\newcommand{\si}{\sigma}

\newcommand{\w}{\wedge}
\newcommand{\wt}{\widetilde}

\newcommand{\ol}{\overline}
\newcommand{\dr}{\partial}

\newcommand{\ap}{\approx}

\let\ssection=\section
\renewcommand{\section}{\setcounter{equation}{0}\ssection}

\newcounter{eqalph}
\newcounter{equationa}
\newcounter{remark}
\newcounter{example}
\newcounter{theorem}
\newcounter{proposition}
\newcounter{lemma}
\newcounter{corollary}
\newcounter{definition}
\newenvironment{eqalph}{\stepcounter{equation}
\setcounter{equationa}{\value{equation}}
\setcounter{equation}{0}

\begin{eqnarray}}{\end{eqnarray}\setcounter{equation}{\value{equationa}}}

\def\theremark{\arabic{remark}}
\def\therexample{\arabic{remark}}

\def\thedefinition{\arabic{definition}}

\newenvironment{theo}{\refstepcounter{definition} \medskip
\noindent{\it Theorem \thedefinition:}}{\medskip}

\newcommand{\mar}[1]{}

\begin{document}
\hbox{}

{\parindent=0pt

{\Large \sc Jacobi fields of completely integrable Hamiltonian systems}

\bigskip

{\sc G.Giachetta\footnote{E-mail: giovanni.giachetta@unicam.it},
L.Mangiarotti\footnote{E-mail: luigi.mangiarotti@unicam.it}} 
\medskip

{\it Department of Mathematics and Physics, University of Camerino,
62032 Camerino (MC), Italy}
\bigskip

{\sc G.Sardanashvily\footnote{ E-mail: sard@grav.phys.msu.su;
URL: http://webcenter.ru/$\sim$sardan/}}

\medskip

{\it Department of Theoretical Physics,
Moscow State University, 117234 Moscow, Russia}
\bigskip
\bigskip

{\small
{\bf Abstract.}

We show that Jacobi fields of a completely integrable Hamiltonian system
of $m$ degrees of freedom also make up a completely 
integrable system. They provide $m$ additional first integrals
which characterize a relative motion.
\medskip

}
}

\bigskip
\bigskip

Given a completely integrable Hamiltonian system (henceforth CIS),
 derivatives of its first integrals
need not be constant on trajectories of a motion. We show that Jacobi fields of 
a CIS provide linear combinations of derivatives of first integrals
which are new integrals of motion which can characterize a relative motion.

Let us consider a Hamiltonian system on a $2m$-dimensional 
symplectic manifold $M$, coordinated by $(x^\la)$ and endowed with a symplectic form
\mar{u1}\beq
\Om=\frac12\Om_{\m\nu}dx^\m\w dx^\nu. \label{u1}
\eeq
The corresponding Poisson bracket reads
\mar{u2}\beq
\{f,f'\}=w^{\al\bt}\dr_\al f\dr_\bt f', \qquad f,f'\in C^\infty(M),\label{u2}
\eeq
where 
\mar{u3}\beq
w=\frac12w^{\al\bt}\dr_\al\w \dr_\bt, \qquad \Om_{\m\nu}w^{\m\bt}=\dl^\bt_\nu,
\label{u3}
\eeq
is the Poisson bivector associated to $\Om$. 
Let a (real smooth) function $H\in C^\infty(M)$ on $M$
be a Hamiltonian of a system in question. Its Hamiltonian vector field
\mar{u4}\beq
\vt_H=-w\lfloor dH=w^{\m\nu}\dr_\m H\dr_v \label{u4}
\eeq
defines the first order Hamilton equation
\mar{u5}\beq
d_t x^\nu=\vt_H^\nu= w^{\m\nu}\dr_\m H \label{u5}
\eeq
on $M$. With respect to the local Darboux coordinates $(q^i,p_i)$,
the expressions (\ref{u1}) -- (\ref{u4}) read
\be
&& \Om=dp_i\w dq^i, \qquad w= \dr^i\w\dr_i, \\
&& \{f,f'\}=\dr^if\dr_if'-\dr_if\dr^i f', \\
&& \vt_H= \dr^if\dr_i -\dr_i f\dr^i.
\ee
The Hamilton equation (\ref{u5}) takes the form
\mar{u6}\beq
d_t q^i=\dr^iH, \qquad d_t p_i=-\dr_iH. \label{u6}
\eeq

A Hamiltonian system $(M,\Om,H)$ is called completely integrable
if there exist $m$ independent first integrals $F_k$ in involution with 
respect to the Poisson bracket (\ref{u2}). Namely, (i) $\{H,F_k\}=0$,
(ii) $\{F_k,F_r\}=0$, and (iii) the differentials $dF_k$ are linearly independent
almost everywhere, i.e., the set of points where this condition fails is nowhere dense
in $M$ \cite{arn,laz}. Of course, a Hamiltonian $H$ itself is a first integral,
but it is not independent of $F_k$. Moreover, one often put $F_1=H$.

Let us consider Jacobi fields of the above CIS
\mar{u20}\beq
(M,\Om,H,F_k). \label{u20}
\eeq
 They are defined as follows \cite{book98,jmp99}. 

Let $TM$ be the
tangent bundle of the manifold $M$ provided with the induced bundle
coordinates $(x^\la,\dot x^\la)$ possessing the transition functions
\be
\dot x'^\la=\frac{\dr x'^\la}{\dr x^\bt}\dot x^\bt.
\ee
Any exterior form
\be
\si=\frac{1}{r!}\si_{\la_1\cdots\la_r}dx^{\la_1}\w\cdots\w dx^{\la_r}
\ee
on $M$ gives rise to the exterior form 
\mar{u7}\beq
\wt\si =\frac1{r!}[\dot x^\m\dr_\m
\si_{\la_1\cdots\la_r}dx^{\la_1}\w\cdots\w dx^{\la_r}+ \op\sum_{i=1}^r
\si_{\la_1\cdots\la_r}dx^{\la_1}\w\cdots\w 
d\dot x^{\la_i}\w\cdots\w dx^{\la_r}] \label{u7}
\eeq
on $TM$ such that the equality
\mar{u10}\beq
d\wt\si=\wt{d\si} \label{u10}
\eeq
holds \cite{leon,grab}. 

In particular, the 
tangent lift (\ref{u7}) of a function $f$ is
$\wt f=\dot x^\la\dr_\la f$. 
The symplectic form $\Om$ (\ref{u1}) on $M$ gives rise to the 2-form
\mar{gm67}\beq
\wt\Om=\frac12(\dot x^\la \dr_\la\Om_{\m\nu}dx^\m\w dx^\nu
+\Om_{\m\nu}d\dot x^\m\w dx^\nu + \Om_{\m\nu}dx^\m\w d\dot x^\nu) \label{gm67}
\eeq
on $TM$. Due to the condition (\ref{u10}), it is a closed form. Written with respect to
the local Darboux coordinates $(q^i,p_i)$ on $M$ and the induced bundle coordinates
$(q^i,p_i,\dot q^i,\dot p_i)$ on $TM$, the form (\ref{gm67}) reads
\mar{u12}\beq
\wt\Om= dp_i\w d\dot q^i + d\dot p_i\w dq^i. \label{u12} 
\eeq
A glance at this expression shows that $\wt\Om$ is a non-degenerate 2-form,
i.e., it is a symplectic form. Note that
the conjugate pairs of coordinates and momenta with respect to this symplectic form are 
$(q^i,\dot p_i)$ and $(\dot q^i,p_i)$. The associated Poisson bracket
on $TM$ reads
\mar{u14}\beq
\{g,g'\}_T=\dr^ig\dot\dr_ig' -\dot\dr_ig\dr^i g' + 
\dot\dr^ig\dr_ig' -\dr_ig\dot\dr^i g', 
\qquad g,g'\in C^\infty(TM), \label{u14}
\eeq
where we have employed the notation
\be
\dot\dr^i=\frac{\dr}{\dr\dot p_i}, \qquad \dot\dr_i=\frac{\dr}{\dr\dot q^i}.
\ee
With the tangent lift 
\mar{u13}\beq
\wt H= \dr_TH, \qquad \dr_T=(\dot q^j\dr_j +\dot p_j\dr^j), \label{u13}
\eeq 
of a Hamiltonian $H$, we obtain a Hamiltonian system $(TM,\wt\Om,\wt H)$ on the
tangent bundle $TM$ of $M$. Computing the Hamiltonian vector field of 
this tangent Hamiltonian with respect to the 
Poisson bracket (\ref{u14}), we obtain
the corresponding Hamilton equations
\mar{u17,8}\ben
&& d_t q^i=\dot\dr^i\wt H=\dr^iH, \qquad d_t p_i=-\dot\dr_i\wt H=-\dr_iH, \label{u17}\\
&& d_t \dot q^i=\dr^i\wt H= \dr_T\dr^iH,
\qquad d_t\dot p_i=-\dr_i\wt H= - \dr_T\dr_iH. \label{u18}
\een
The equation (\ref{u17}) coincides with the Hamilton equation (\ref{u6}) of the
original Hamiltonian system on $M$, while the equation (\ref{u18}) 
is the well-known variation equation of the equation (\ref{u17}).
Substituting a solution $s$ of the Hamilton equation (\ref{u17}) into (\ref{u18}),
one obtains a linear dynamic equation whose solutions $\ol s$ are the
Jacobi fields of the solution $s$. Indeed, if $M$ is a vector space,
there is the canonical splitting $TM\ap M\times M$, and  
$s+\ol s$ is a solution of the Hamilton equation 
(\ref{u17}) modulo terms of order $>1$ in $\ol s$.

Note that one can define Jacoby fields of Euler--Lagrange equations 
in a similar way \cite{book98,book,nun}.

Turn now to integrals of motion of the Hamiltonian system 
$(\wt\Om,\wt H)$ on $TM$. We will denote the pull-back onto $TM$ of a 
function $f$ on $M$ by the same symbol $f$.
The Poisson bracket $\{.,.\}_T$ (\ref{u14})
possesses the following property. Given arbitrary functions
$f$ and $f'$ on $M$ and their tangent lifts $\dr_Tf$ and $\dr_Tf'$
on $TM$, we have the relations
\mar{u19}\beq
\{f,f'\}_T=0, \quad \{\dr_T f,f'\}_T=\{f,\dr_Tf'\}_T=\{f,f'\}, \quad
\{\dr_T f,\dr_T f'\}=\dr_T\{f,f'\}. \label{u19}
\eeq
Let us consider the tangent lifts $\dr_TF_k$ of first integrals $F_k$ 
of the original CIS (\ref{u20}) on $M$. By virtue of the relations (\ref{u19}), 
the functions $(F_k,\dr_T F_k)$
make up a collection of $2m$ first integrals of the tangent 
Hamiltonian system $(\wt\Om,\wt H)$ on $TM$, i.e., they are constant on
solutions of the Hamilton equations (\ref{u17}) -- (\ref{u18}). 
It is readily observed that these first integrals are independent on $TM$.
Consequently, we have a CIS
\mar{u21}\beq
(TM,\wt\Om,\wt H,F_k,\dr_T F_k) \label{u21}
\eeq
on the tangent bundle $TM$.

Since the first integrals $\dr_T F_k$ 
of the CIS (\ref{u21}) depend on Jacobi fields, one may hope 
that they 
characterize a relative motion. Given a solution
$s(t)$ of the Hamilton equation (\ref{u6}), one could approximate 
other solutions
$s'(t)$ with initial data $s'(0)$ close to $s(0)$ are approximated
$s'\ap s+\ol s$ by solutions $(s,\ol s)$ of the Hamilton equations 
(\ref{u17}) -- (\ref{u18}). However, such an approximation need not be
good. Namely, if $M$ is a vector space and $s'(0)=s(0) +\ol s(0)$
are the above mentioned solutions, 
the difference $s'(t)-(s(t) +\ol s(t))$, $t\in\Bbb R$,
fails to be zero and, moreover, need not be bounded in $M$.  
Of course, if $F_k$ is a first integral, then
\be
F_k(s'(t))- F_k(s(t))={\rm const.}
\ee
whenever $s$ and $s'$ are solutions of the 
Hamilton equation (\ref{u6}). We aim to show that, under a certain condition,
there exists a Jacobi field $\ol s$ of the solution $s$ such that
\mar{u22}\beq
F_k(s')- F_k(s)=\dr_TF_k(s,\ol s) \label{u22}
\eeq
for all first integrals $F_k$ of the CIS (\ref{u20}). 
It follows that, given a trajectory $s$ of the original CIS (\ref{u20})
and the values of its first integrals $F_k$ on $s$, we can restore the values
of $F_k$ on other trajectories $s'$ from $F_k(s)$ and the values of 
first integrals $\dr_TF_k$ for different Jacobi fields of the solution $s$.
Therefore, one may say that the first integrals $\dr_TF_k$ of the tangent CIS 
(\ref{u21}) characterize a relative motion.

We refer to the following 
assertion \cite{theo}.

\begin{theo} \label{z8} \mar{z8}
Let $N$ be a connected invariant manifold of the CIS
 (\ref{u20}). Let $U$ be an open neighbourhood of $N$ 
such that: (i) first integrals $F_k$ are independent
everywhere in $U$, (ii) the Hamiltonian vector fields of 
the first integrals $F_k$ on $U$ are complete,
and (iii) the submersion $\times F_k: U\to \Bbb R^m$
is a trivial bundle of invariant manifolds 
over a domain $V\subset \Bbb R^m$. 
Then $U$ is isomorphic
to the symplectic annulus 
\mar{z10}\beq
W=\Bbb R^{m-n}\times T^n\times V, \label{z10}
\eeq
provided with the action-angle coordinates 
\mar{u23}\beq
(z^1,\ldots, z^{m-n}; z^{m-n+1},\ldots,z^m; I_1,\ldots,I_m) \label{u23}
\eeq
where $(z^{m-n+1},\ldots,z^m)$ are cyclic coordinates on the torus $T^n$.
Written with respect to these coordinates, the symplectic form on $W$ reads
\be
\Om=dI_i\w dz^i, 
\ee
while a Hamiltonian $H$ and the first integrals $F_k$ depend only on  
action coordinates $I_i$.
\end{theo}

Note that, if $N$ is compact, the conditions (ii) and (iii) of Theorem
\ref{z8} always hold, and we come to the classical Arnold--Liouville theorem
\cite{arn}.

Let $N$ and $U$ be as in Theorem \ref{z8}. Let us 
consider the restriction of the CIS
(\ref{u20}) to $U$. Passing to the action-angle coordinates (\ref{u23}),
we obtain an equivalent CIS on the symplectic 
annulus $W$ (\ref{z10}). The Hamilton equation (\ref{u6}) on $W$ reads
\mar{u24}\beq
d_t z^i=\dr^i H(I_j), \qquad d_t I_i=0. \label{u24}
\eeq
Let us consider the tangent CIS on the tangent bundle $TU$ of $U$.
It is the restriction to $TU\subset TM$ of the tangent CIS (\ref{u21}) on $TM$. 
Equipping $TU$ with the induced bundle coordinates
\mar{u25}\beq
(z^i, I_i,\dot z^i,\dot I_i), \label{u25}
\eeq
we obtain the tangent CIS on $TM$.
The tangent Hamiltonian form $\wt\Om$
on $TW$ reads
\be
\wt\Om=dI_i\w d\dot z^i +d\dot I_i\w d z^i. 
\ee
Its Hamiltonian 
\be
\wt H=\dr_TH=\dot I_i\dr^i H
\ee
and first integrals $(F_k,\dr_T F_k)$ depend only on coordinates $(I_j,\dot I_j)$. Thus, the 
coordinates (\ref{u25}) are the action-angle coordinates on the symplectic annulus
\be
TW=V\times \Bbb R^{3m-n}\times T^n.
\ee
The Hamilton equations (\ref{u17}) -- (\ref{u18}) on $TW$ read
\mar{u26,7}\ben
&& d_t I_i=0, \qquad d_t \dot I_i=0, \label{u26}\\
&& d_t z^i=\dr^i H(I_j), \qquad d_t \dot z^i=\dot I_k\dr^k\dr^i H(I_j). \label{u27}
\een

Let $s$ and $s'$ be solutions of the Hamilton equation (\ref{u6}) which live
in $U$. Consequently, they are solutions of the Hamilton equation
(\ref{u24}) on $W$. Hence, their action
components
$s_i$ and $s'_i$ are constant. Let us consider the system of algebraic
equations
\be
F_k(s'_j)-F_k(s_j)=a_i\dr^iF_k(s_j), \qquad k=1,\ldots,m,
\ee
for real numbers $a_i$, $i=1,\ldots m$.
Since the first integrals $F_k$ have no critical points on $W$,
this system always has a unique solution. Then let us choose a solution
$(s,\ol s)$ of the Hamilton equations (\ref{u26}) -- (\ref{u27}), where the 
Jacobi field $\ol s$ of the solution $s$ possess the action components
$\ol s_i=a_i$. It fulfills the relations (\ref{u22}) 
for all first integrals $F_k$. 
In other words, first integrals $(F_k,\dr_T F_k)$ on $TU$ can be replaced by
the action variables $(I_k,\dot I_k)$. Given a solution $I$ of the Hamilton equation
(\ref{u24}), its another solution $I'$ is approximated good by the solution
$(I,\dot I=I'-I)$ of the Hamilton equation (\ref{u26}).

\end{document}